\documentstyle[12pt]{article}
\def\journal#1, #2, #3, #4 { {\sl #1~}{\bf #2~}(#3) #4 }

\def\cmp{\journal Comm. Math. Phys., }

\def\np{\journal Nucl. Phys., }

\def\pl{\journal Phys. Lett., }

\catcode`\@=11
\def\marginnote#1{}
\newcount\hour
\newcount\minute
\newtoks\amorpm
\hour=\time\divide\hour by60
\minute=\time{\multiply\hour by60 \global\advance\minute
by-\hour}\edef\standardtime{{\ifnum\hour<12
\global\amorpm={am}%
        \else\global\amorpm={pm}\advance\hour by-12 \fi
        \ifnum\hour=0 \hour=12 \fi
        \number\hour:\ifnum\minute<10
0\fi\number\minute\the\amorpm}}
\edef\militarytime{\number\hour:\ifnum\minute<10
0\fi\number\minute}

\def\draftlabel#1{{\@bsphack\if@filesw {\let\thepage\relax
   \xdef\@gtempa{\write\@auxout{\string
      \newlabel{#1}{{\@currentlabel}{\thepage}}}}}\@gtempa
   \if@nobreak \ifvmode\nobreak\fi\fi\fi\@esphack}
        \gdef\@eqnlabel{#1}}
\def\@eqnlabel{}
\def\@vacuum{}
\def\draftmarginnote#1{\marginpar{\raggedright\scriptsize\tt#1}}\def\draft{\oddsidemargin -.5truein
        \def\@oddfoot{\sl preliminary draft \hfil
        \rm\thepage\hfil\sl\today\quad\militarytime}
        \let\@evenfoot\@oddfoot \overfullrule 3pt
        \let\label=\draftlabel
        \let\marginnote=\draftmarginnote

\def\@eqnnum{(\theequation)\rlap{\kern\marginparsep\tt\@eqnlabel}%
\global\let\@eqnlabel\@vacuum}  }

\def\numberbysection{\@addtoreset{equation}{section}
        \def\theequation{\thesection.\arabic{equation}}}

\def\underline#1{\relax\ifmmode\@@underline#1\else
 $\@@underline{\hbox{#1}}$\relax\fi}

\catcode`@=12
\relax

\numberbysection
\def\souligne#1{\underline{#1}}
\def\beq{\begin{equation}}
\def\eeq{\end{equation}}
\def\beqa{\begin{eqnarray}}
\def\eeqa{\end{eqnarray}}

\def\sqr#1#2{{\vcenter{\vbox{\hrule height.#2pt
\hbox{\vrule width.#2pt height#1pt \kern#1pt
\vrule width.#2pt}
\hrule height.#2pt}}}}

\def\hhat{{\widehat h}}

\def\Jhat{{\widehat J}}

\def\xihat{{\widehat \xi}}

\def\xib{{\overline \xi}}

\def\qhat{{\widehat q}}

\def\zb{{\overline z}}

\def\Jb{{\bar J}}

\def\rhob{\bar \rho}

\def\CP{{\cal CP}^n}

\def\Xb{{\bar X}}
\def\Yb{{\bar Y}}
\def\Ab{\bar A}

\def\Ub{{\bar U}}
\def\zb{\bar z}
\def\ab{\bar a}

\def\eb{\bar e}
\def\fb{\bar f}

\def\jb{\bar \jmath}

\def\betab{\bar \beta}
\def\chib{\bar \chi}

\def\1b{\bar 1}

\def\Wb{\bar W}

\def\xib{\bar \xi}
\def\mub{\bar \mu}

\def\bff{{\bf f}}
\def\bbff{\bar {\bf f}}
\def\bfe{{\bf e}}
\def\bbfe{\bar {\bf e}}
\def\bbfe{\bar {\bf e}}
\def\bfX{{\bf X}}

\def\partialb{\bar \partial}

\def\0b{\bar 0}

\def\Yb{\bar Y}
\def\Jb{\bar J}

\def\rhob{\bar{\rho}}
\def\psis{\psi^+}

\def\bra#1{\langle {#1}|}
\def\ket#1{|{#1}\rangle}
\def\psif#1{\psi_{\displaystyle{f^{#1}}}}
\def\psifs#1{\psi^+_{\displaystyle{\fb^{#1}}}}
\def\psifz#1#2{\psi_{\displaystyle{f^{#1} }({#2})}}
\def\psifsz#1#2{\psi^+_{\displaystyle{\fb^{#1}}({#2})}}
\def\eJz{e^{\sum_0^n J_sz^{(s)}}}
\def\eJzb{e^{\sum_0^n \bar{J}_t\bar{z}^{(t)}}}

\def\omegab{\bar \omega}

\def\betab{\bar \beta}

\def\Jb{\bar J}

\def\omegab{\bar \omega}

\def\eb{\bar e}
\def\varphib{\bar \varphi}

\newtheorem{definition}{\bf Definition}
\newtheorem{theorem}{\bf Theorem}

\newtheorem{proposition}{\bf Proposition}

 \font\tenrm=cmr10

\begin{document}
\renewenvironment{thebibliography}[1]
  { \begin{list}{\arabic{enumi}.}
    {\usecounter{enumi} \setlength{\parsep}{0pt}
     \setlength{\itemsep}{3pt} \settowidth{\labelwidth}{#1.}
     \sloppy
    }}{\end{list}}

\parindent=1.5pc
\begin{flushright}

hep-th/9212109
\\
LPTENS--92/36
\\
December 1992
\end{flushright}
\vglue 1 true cm

\begin{center}{\bf RECENT PROGRESS OF  \\
               \vglue 3pt
               THE  LIOUVILLE  APPROACH TO 2D GRAVITY   \\
               \vglue 3pt
              AND ITS TODA (W) GENERALIZATIONS                }\\
\vglue 1.0cm
{ JEAN-LOUP~GERVAIS}\\
\baselineskip=14pt
{\it  Laboratoire de Physique Th\'eorique de
l'\'Ecole Normale Sup\'erieure\footnote{Unit\'e Propre du
Centre National de la Recherche Scientifique,
associ\'ee \`a l'\'Ecole Normale Sup\'erieure et \`a l'Universit\'e
de Paris-Sud.},\\
24 rue Lhomond, 75231 Paris CEDEX 05, ~France.}
\vglue .8 true cm
{ABSTRACT}
\end{center}
\vglue 0.3cm
{\rightskip=3pc
 \leftskip=3pc
 \tenrm\baselineskip=12pt
 \noindent
These lecture notes review current  progress on the class of
conformal theories which may be studied by quantizing the
conformal Toda dynamics.  After summarizing recent developments of the
quantum Liouville theory,  one recalls how two-dimensional
black holes come out from the
non-abelian Toda systems, and reviews the geometrical interpretation of
the $A_n$-Toda theories,  just put forward, that relate W geometries
with the external geometry of particular (W) surfaces.

\vglue 0.8cm}

\section{Introduction}
The connection between conformally invariant field theories
(CIFT) and integrable models has proven to be more
and more useful in recent years. A particular role
is played in this connection by the conformal Toda theories
that generate, upon quantization, a whole class of the
most important CIFT's. Two-dimensional gravity in
the conformal gauge is
notoriously equivalent to  the Liouville theory which is
the conformal Toda theory
 associated with the Lie algebra $A_1$.
A step to generalize this situation was made in \cite{BG1} where it was
shown that the  Toda theory associated\footnote{Except in section 3,
we only consider the so-called principal grading (see section 3)}
 with any given simple
Lie algebra gives  two Noether realizations of the corresponding
W-algebra.
Thus, if the above $A_1$-Liouville scheme
is repeated for the other Lie algebras, there should exist
generalizations
of two-dimensional gravity (called W-gravities) which
are invariant by
generalized diffeomorphisms, and coincide with
the conformal Toda theories
when a particular local coordinate frame is used.

These lecture notes cover recent developments.

In \souligne{section 2}, Liouville theory is discussed using the
operator approach\cite{GN4,G1,G3,G4,G5}, where the quantum-group structure
plays  a fundamental role. After recalling some background material,
one reviews recent results:

\noindent 1) The complete derivation\cite{CGR}  of the holomorphic
operator-algebra
(braided category) where the role of the quantum group structure
is fully established. The main new result here is that the fusion and
braiding are not completely given by the quantum-group
3-j or 6-j symbols, as is usually assumed: There are coupling
constants, which are not trigonometrical functions, and are
not determined solely by the quantum-group symmetry.

\noindent 2) The derivation\cite{G5}  of the
gravity-matter coupling in the weak-coupling regime, where the
present method is found to agree with matrix-model calculations.
One crucial point here is to show that the continuation
in the number of screening operators which  is made in the
Coulomb gas picture is equivalent to  the symmetry between
quantum-group spin $J$ and $-J-1$ put forward in my  previous
study\cite{G2,G3} of the strong-coupling regime.

The remaining sections cover recent progress about the more
general Toda theories, considered at the purely
classical level.

In \souligne{section 3}, it is first recalled how Toda theories
are associated with any given embedding of $sl(2)$ into a given
simple algebra. This defines a gradation.
If the subgroub of gradation zero is non-abelian, the Toda
theory is called non-abelian. The point of section 3 is to
review the recent discovery\cite{GS}  of the black-hole background
metric of these
non-abelian Toda theories.

The basic point of \souligne{section 4}
is the recent proposal\cite{GM1,GM2}
that one can regard the W-geometries  as
the extrinsic geometry of particular  two dimensional surfaces
(W-surface) embedded into target spaces that are
  higher dimensional K\"ahler manifolds
(We will restrict ourselves to the simplest   particular situation,
i.e. our target space is $\CP$ which corresponds to the $A_n$-type
W-geometry).
Instead of introducing higher-spin gauge generators,
our approach makes use of the {\it extrinsic} curvatures of
the embedded surface at its regular points,
and relates  it with the Toda dynamics mentioned
above.  The main virtue of our approach is that it is very simple
to begin with.
A W surface is characterized by the specific
{\it chiral structure} of its
embedding which we
 call chiral for short.

\souligne{Section 5} deals with the geometry of the Toda hierachy.
The dynamical variables  of this hierarchy are shown to give  particular
coordinates  of the higher dimensional K\"ahler manifolds.
They are delt with by means of the free-fermion formalism,
which is shown to be deeply connected with the concept of
analytic continuation and with its W-generalizations.
An important  point here is to show that this allows us to extend the
W transformations  as
reparametrizations  of the target space, so that they become
linear.

Finally, in  \souligne{section 6}  we first
reformulate our
approach in terms of the intrinsic geometry of the family of
associated surfaces in the Grassmannians $G_{n+1, k+1}$,
$k=1,\, \cdots,\, n$. This is needed to study  singular points
 and global aspects of W-surfaces following the
general scheme of refs. \cite{GM1,GM2}, where
these surfaces were shown to be  intanton solutions of the associated
non-linear $\sigma$ models.
The aim is to establish the
generalization of the Gauss-Bonnet theorem to the W-surfaces
discussed above.  The instanton-number associated with each mapping
are  the global invariant of W-geometry.
They are connected  with the singularity indices of the W-surface.

\section{Two dimensional gravity}
\subsection{The classical structure}
let us first  recall some feature of the
classical  Liouville dynamics. In the conformal gauge,
it  is governed by the action:
\begin{equation}
 {\it S}={1\over 4\pi } \int d_2x
\sqrt{\widehat g} \Bigl\{ {1\over 2} {\widehat
g}^{ab}\partial_a\Phi
\partial_b\Phi+
e^{\displaystyle 2\sqrt{\gamma} \Phi}
+{1\over 2\sqrt{\gamma}}R_0 \Phi\Bigr \}
\label{2.1}
\end{equation}
 ${\widehat g}_{ab}$ is the fixed background metric. We work
for fixed genus, and do not integrate over the moduli. As  is
well known,  one can choose a local coordinate system such that
${\widehat
g}_{ab}=\delta_{a b}$.
 Thus we are reduced to the action
\begin{equation}
 {\it S}={1\over 4\pi } \int d\sigma d\tau
\, \Bigl ( {1\over 2}({\partial \Phi\over \partial \sigma})^2
+{1\over 2}({\partial \Phi\over \partial \tau})^2
+e^{\displaystyle 2\sqrt{\gamma} \Phi} \Bigr )
\label{2.2}
\end{equation}
where $\sigma$ and $\tau$ are the local coordinates.  The
 complex structure is assumed to be such that the curves
with
constant $\sigma$ and $\tau$ are everywhere
tangent to the local imaginary
and real axis  respectively.
In a typical situation, one may  work on the cylinder $0\leq
\sigma \leq 2\pi$,  $-\infty \leq \tau \leq \infty $ obtained by
an
appropriate mapping from one of the handles of a general
Riemann surface, and we shall do so in the present article.
The action \ref{2.2}  corresponds to a conformal
 theory such that $\exp(2\sqrt{\gamma} \Phi) d\sigma d\tau $
is invariant.
  The classical equivalent of the chiral vertex operators
 may be obtained  very simply\cite{GN1,GN2,G4} by   using the fact that
the field $\Phi (\sigma,\, \tau)$
satisfies the equation
\begin{equation}
 {\partial^2 \Phi\over \partial \sigma^2}
+{\partial^2 \Phi\over \partial \tau^2}=
2 \sqrt{\gamma}\> e^{\displaystyle 2\sqrt{\gamma}\Phi}
\label{2.3}
\end{equation}
if and only  if
\begin{equation}
e^{-\displaystyle \sqrt{\gamma} \Phi}={i\sqrt{\gamma \over 2}}
\sum_{j=1,2} f_j(x_+)
g_j(x_-);
  \quad  x_\pm=\sigma\mp i\tau
\label{2.4}
\end{equation}
where $f_j$ (resp.($g_j$), which are functions of a single
variable,  are
solutions of the  same Schr\"odinger equation
\begin{equation}
-f_j''+T(x_+)f_j=0,\quad
\hbox{( resp.}\>  -g''_j+\overline T(x_-)g_j\,\hbox{)}.
\label{2.5}
\end{equation}
 The solutions
are normalized so that their Wronskians $f_1'f_2-f_1f_2'$and $g_1'g_2-g_1g_2'$
are equal to one.
The proof of this basic fact is straightforward
\cite{GN1,GN2,G4}. The potentials
$T(x_+)$ and $\overline T(x_-)$ are the two
components of the stress-energy tensor, and, after quantization,
Eqs.\ref{2.5}  become the  Virasoro Ward-identities
 associated with  the vanishing of the singular vector
at the second level. As a result the Liouville theory also
describes minimal models provided the coupling constant
$\gamma$ is taken to be negative. This is how, we shall treat
the matter fields.
 For the dynamics associated with the
action Eq.\ref{2.2}, $\tau$ is the time variable, and the canonical
Poisson
brackets are
\begin{equation}
\bigl \{\Phi(\sigma_1, \tau),
{\partial\over \partial \tau}  \Phi(\sigma_2, \tau) \bigr\}_{\hbox{P.B.}}
=4\pi \,  \delta(\sigma_1-\sigma_2),\quad
\bigl \{\Phi(\sigma_1, \tau),\Phi(\sigma_2, \tau) \bigr
\}_{\hbox{P.B.}}= 0
\label{2.6}
\end{equation}
The cylinder  $0\leq
\sigma \leq 2\pi$,  $-\infty \leq \tau \leq \infty$
 may be mapped on the complex
plane of $z=e^{\tau+i\sigma}$, and the above Poisson brackets
lead to the usual radial quantization.

A priori, any two
pairs
$f_j$ and $g_j$ of linearly independent solutions of Eq.\ref{2.5}
are suitable. In this connection, it is convenient to rename the
functions $g_j$ by letting $\overline f_1=- g_2$,
$\overline f_2= g_1$.
 Then   one  easily sees  that Eq.\ref{2.4} is
left unchanged if $f_j$ and $\overline f_j$ are replaced
by
$\sum_kM_{jk}f_k$ and   $\sum_kM_{kj}\overline f_k$,
respectively, where
$M_{jk}$ is an arbitrary constant matrix with
 determinant  equal to one.
Eq.\ref{2.4} is
$sl(2,C)$-invariant with $f_j$ transforming as a representation of
spin $1/2$. At the quantum level, the $f_j$'s and $\overline f_j$'s
 become operators that do not commute, and  the
group $sl(2)$ is deformed to become  the quantum group
$U_q(sl(2))$.
 This  structure plays a crucial role at
 the quantum level, and we now elaborate
upon  the classical  $sl(2)$ structure  where the calculations are simple.

At the classical level, it is trivial to take Eq.\ref{2.4} to any
power.  For positive integer  powers $2J$, one gets
(letting $\beta=i\sqrt {\gamma \over 2}$)
\beq
e^{-\displaystyle 2J \sqrt{\gamma} \Phi}=
\sum_{M=-J}^J {\beta^{2J}(-1)^{J-M} (2J) !\over (J+M) !(J-M) !}
\left (f_1(x_+)\,\overline f_2(x_-)\right )^{J-M}
\left(f_2(x_+) \overline f_1(x_-)\right )^{J+M}.
\label{2.7}
\eeq
It is convenient  to put the result under the form
\begin{equation}
e^{-\displaystyle 2J \sqrt{\gamma} \Phi}=
\beta^{2J}
\sum_{M=-J}^J (-1)^{J-M}
f_M^{(J)}(x_+) \overline f_{-M}^{(J)}(x_-).
\label{2.8}
\end{equation}
where $J\pm M$ run over integer. The $sl(2)$-structure has been  made
transparent by letting
\beq
f_M^{(J)}\equiv \sqrt { \textstyle {2J\choose J+M}}
\left (f_1\right )^{J-M} \left(f_2\right )^{J+M},  \quad
\overline f_M^{(J)}\equiv\sqrt { \textstyle {2J\choose J+M}}
\left (\overline f_1\right )^{J+M}
\left(\overline f_2\right )^{J-M}.
\label{2.9}
\eeq
The notation anticipates that $f_M^{(J)}$ and $\overline
f_M^{(J)}$
form   representations of spin $J$. This is  indeed true
since
$f_1$, $f_2$ and $\overline f_1$, $\overline f_2$    span
spin $1/2$ representations, by construction.   Explicitely one finds
$$I_\pm f_M^{(J)}=\sqrt {(J\mp M)(J\pm M+1)} f_{M\pm 1}^{(J)},
\quad
I_3 f_M^{(J)} =Mf_M^{(J)}
$$
\begin{equation}
\overline I_\pm \overline f_M^{(J)}
=\sqrt {(J\mp M)(J\pm M+1)} \overline f_{M\pm 1}^{(J)}, \quad
\overline I_3 f_M^{(J)} =M\overline f_M^{(J)},
\label{2.10}
\end{equation}
where $ I_\ell$ and $\overline I_\ell$ are the
infinitesimal generators of the   $x_+$ and  $x_-$ components
respectively. Moreover, one sees that
\begin{equation}
\left (I_\ell+ \overline I_\ell\right )
e^{-\displaystyle 2J \sqrt{\gamma} \Phi}=0
\label{2.11}
\end{equation}
so that the exponential of the Liouville field are group
invariants.

\subsection{The basic chiral operator-algebras}

Denote by $C$ the central charge of gravity. The standard screening
charges
$-\alpha_\pm$ of the Liouville theory\cite{GN4,GN6} are such that
$$
\alpha_\pm={1\over 2} \Bigl (\sqrt{ {C-1\over 3}}
\pm \sqrt{ {C-25\over 3}} \Bigr),
$$
\begin{equation}
\alpha_\pm={Q\over 2}\pm \alpha_0, \quad
Q=\sqrt{ {C-1\over 3}}, \quad
\alpha_0={1\over 2} \sqrt{ {C-25\over 3}}
\label{2.12}
\end{equation}
$Q$, and $\alpha_0$ are introduced so that they  agree with the
standard notations.
Kac's formula
  may be written as
\begin{equation}
 \Delta_{Kac} (J,\Jhat;C)=-{1\over 2} \beta(J,\Jhat;C)
\Bigl (\beta(J,\Jhat;C)+Q\Bigr), \quad
\beta(J,\Jhat;C)=J\alpha_-+\Jhat\alpha_+,
\label{2.13}
\end{equation}
where $2J$ and $2\Jhat$ are positive integers.
We shall deal with generic values of $C$ in order to avoid
the complications of quantum group representations.
As a result, we have to deal with  $h$ and $\hhat$ on the
same basis. This is in fact crucial for dealing with the
strong-coupling regime\cite{GN6,G2,G3}.   There are two quantum-group
parameters $h=\pi (\alpha_-)^2/2$, with  $q=e^{ih}$, and
$\hhat=\pi (\alpha_+)^2/2$, with  $\qhat=e^{i\hhat}$.
Thus the number of fields doubles
with respect with the classical case just recalled. The holomorphic
operator will have a hat, or not, if they are related with $\hhat$,
or $h$, respectively; and the antiholomorphic operators
are distinguished by an additional bar. According to Eq.\ref{2.13},
the most general Liouville field is to  be
written  as $\exp \bigl(-(J\alpha_-+\Jhat\alpha_+)\Phi\bigr)$.
These fields have decompositions onto holomorphic and
antiholomorphic  operators to which we shall come below.

First we
concentrate on holomorphic fields
  which are only
functions of $x=\sigma-i\tau$. Then the whole structure may be
described on the unit circle $\tau=0$. The holomorphic operators
associated with $h$ form a subfamily which we next describe
following refs.\cite{G1,G3,CGR}.
There are two useful basis of operators.
First, the  vertex operators $V_{m}^{(J)}$, with  $-J\leq m\leq J$,
and $2J$ a positive integer. They diagonalize the monodromy matrix
\begin{equation}
V_m^{(J)}(\sigma+2\pi ) =  e^{2ihm\varpi}
e^{2ihm^2}\,
V_m^{(J)}(\sigma),
\label{2.14}
\end{equation}
where $\varpi$ is the Liouville quasi-momentum (zero mode)
around the cylinder.
The meaning of the index $m$ is to specify the shift of $\varpi$:
\begin{equation}
V_m^{(J)}(\sigma)\>\varpi=(\varpi+2m)\>V_m^{(J)}(\sigma).
\label{2.15}
\end{equation}
The Liouville momentum has been normalized so that this shift is
integer.
$\varpi$ characterizes the Verma module.    We shall work
 on the sphere, where
the  spectrum of $\varpi$ is  of the form
$\varpi_J=\varpi_0+2J$. The momentum  $\varpi_0=1+2\pi/h$
is the one  of the
$sl(2,C)$ invariant state.  There is a
 Verma modules
${\cal H}_J$ for each $\varpi_J$.  The Moore Seiberg (MS)
chiral vertex-operators connect three specified Verma modules
and are thus  of the form
$\phi_{J_3,J_1}^{J_2}$. The operators $V_m^{(J)}$,
on the contrary,
act in the direct sum
${\cal H}=\oplus_J {\cal H}_J$. It is quite obvious, according to
Eq.\ref{2.15}, that   the two are
related by the projection operator ${\cal P}_{J}$:
\beq
{\cal P}_{J} {\cal H} ={\cal H}_J, \quad
{\cal P}_{J_3} V_m^{(J_2)} \equiv \phi_{J_3,J_3+m}^{J_2}
\label{2.16}
\eeq
 The $V$ fields are such that $<\varpi_2 | V_m^{(J)} |\varpi_1>$ is
equal to
one of $\varpi_1=\varpi_3+2m$, and is equal to zero
otherwize. This normalization is  required by the symmetry between three
legs (sphere with three punctures).
The   complete
fusion and braiding algebras
take the form:
$$
{\cal P}_{K}\> V^{(J_1)}_{m_1}\,V^{(J_2)}_{m_2} =
\sum _{J= \vert J_1 - J_2 \vert} ^{J_1+J_2}
 F_{K+m_1,J}\!\!\left[
^{J_1}_{K}
 \> ^{\quad J_2}_{K+m_1+m_2 }
\right]
$$
\beq
\sum _{\{\nu\}}
{\cal P}_{K}\> V ^{(J,\{\nu\})} _{m_1+m_2}
<\!\varpi_{J},{\{\nu\}} \vert V ^{(J_1)}_{J_2-J}
\vert \varpi_{J_2} \! >
\label{2.17}
\eeq
where ${\cal P}_{K}$ is the projector onto ${\cal H}_{K}$, and,
\beq
{\cal P}_{K}\> V^{(J_1)}_{m_1}\,V^{(J_2)}_{m_2} =
\sum _{n_2}
 B_{K+m_1,K+n_2 }\!\!\left[
^{J_1}_{K}
 \> ^{\quad J_2}_{K+m_1+m_2 }
\right]
{\cal P}_{K}\> V ^{(J_2,)} _{n_2}
 V ^{(J_1)}_{m_1+m_2-n_2}
\label{2.18}
\eeq
For simplicity of notation we omitted the dependence upon the
world-sheet variables, which is standard. Using Eq.\ref{2.16}, one
may verify
that these expressions
have the general MS form.   In ref.\cite{CGR}, it was shown that
\beq
F_{{J_{23}},{J_{12}}}\!\!\left[^{J_1}_{J_{123}}
\,^{J_2}_{J_3}\right]
=
{g_{J_1J_2}^{J_{12}}\
g_{J_{12}J_3}^{J_{123}}
\over
g _{J_2J_3}^{J_{23}}\
g_{{J_1}J_{23}}^{J_{123}}
}
\left\{ ^{J_1}_{J_3}\,^{J_2}_{J_{123}}
\right. \left |^{J_{12}}_{J_{23}}\right\}
\label{2.19}
\eeq
where $\left\{ ^{J_1}_{J_3}\,^{J_2}_{J_{123}}
\right. \left |^{J_{12}}_{J_{23}}\right\}$ denote the 6-j symbols
of $U_q(sl(2))$, with $q=e^{ih}$. This term was of course expected,
in view of the quantum-group structure previously exhibited,
in particular,  in refs.\cite{G1,G3}.
However, there appear, in addition,  coupling constants
$g_{J_1J_2}^{J_{12}}$, which are not trigonometrical functions of $h$.
Their general expression is
\beq
g_{J_1J_2}^J
=
\prod_{n=1}^{J_1+J_2-J}
\sqrt{
{G(J_1-{n/ 2})
G(J_2-{n/ 2})
G(J+{n/ 2})
\over
G({(n-1)/2})}
},
\label{2.20}
\eeq
where
\beq
G(J)\equiv
{\Gamma \left (1+(1+2J)h/\pi \right )\over
\Gamma \left (-(1+2J)h/\pi\right )}.
\label{2.21}
\eeq
$\Gamma(x)$ is the usual (not $q$-deformed) $\Gamma$-function.
The relation between braiding and fusion matrices is standard\cite{MS},
and one may verify that the appearance of the coupling constant does not
spoil the consistency conditions (pentagonal, and so on) satisfied
by the 6-j symbols.
In terms of the $V$ fields the operator-algebra is entirely expressed
in terms of quantum group invariant. Thus, it is clearly symmetric, but
one does not see how the quantum group acts. This last feature is
exhibited\cite{B}
by going to another basis of operators denoted $\xi_M^{(J)}$
 (the quantum group basis) where $M$ will really be the ``third''
 component of the quantum-group spin. 	The change of basis
is of the form\cite{B,G1}
\beq
{\cal P}_K \xi_M^{(J)}(\sigma) := \sum_m C_{K K+2m} (J,M)
\> {\cal P}_K V_m^{(J)}(\sigma).
\label{2.22}
\eeq
where the coefficient $C_{K K+2m} (J,M)$ is proportional to
a q-deformed hypergeometric function.
The fusion properties become\cite{CGR}
\beq
\xi ^{(J_1)}_{M_1}\,\xi^{(J_2)}_{M_2} =
 \sum _{J= \vert J_1 - J_2 \vert} ^{J_1+J_2}
g _{J_1J_2}^J \, (J_1,M_1;J_2,M_2\vert J)
\sum _{\{\nu\}} \xi ^{(J,\{\nu\})} _{M_1+M_2}
<\!\varpi _J,{\{\nu\}} \vert V ^{(J_1)}_{J_2-J}
\vert \varpi_{J_2}\! >,
\label{2.23}
\eeq
where $(J_1,M_1;J_2,M_2\vert J)$ denotes the q-deformed Clebsch-Gordan
(CG) coefficients. The braiding relations
are\footnote{We only coonsider
one of the two possible orderings on the unit circle}
as follows\cite{G1}
\beq
\xi ^{(J_1)}_{M_1}\,\xi^{(J_2)}_{M_2}
=\sum_{N_1 N_2} (J_1,J_2)_{M_1\, M_2}^{N_2\, N_1}\>
\xi^{(J_2)}_{M_2}\,\xi ^{(J_1)}_{M_1}.
\label{2.24}
\eeq
The braiding matrix $(J_1,J_2)_{M_1\, M_2}^{N_2\, N_1}$
is given by the
matrix element of the universal $R$-matrix of $U_q(sl(2))$:
\beq
(J_1,J_2)_{M_1\, M_2}^{N_2\, N_1}=
\Bigl(<\! <J_1,M_1\vert \otimes <\! <J_2,M_2
\vert\Bigr)\> {\bf R}
\>\Bigl(\vert J_1,N_1>\! > \otimes \vert J_2,N_2>\! >\Bigr).
\label{2.25}
\eeq
$|J,M>\! >$ are group theoretic states which span the
representation of spin $J$ of $U_q(sl(2))$, and
 $R$ is the universal R-matrix:
\beq
{\bf R}= e^{-2ihJ_3 \otimes J_3}
\sum_{n=0}^\infty \,
{{(1-e^{2ih})^{n}\,e^{ihn(n-1)/2} \over
\lfloor n \rfloor \! !}} e^{-ihnJ_3}(J_+)^n \otimes
e^{ihnJ_3}(J_-)^n,
\label{2.26}
\eeq
 $J_\pm$, and $J_3$ are the quantum-group generators, which satisfy
\begin{equation}
\Bigl[J_+,J_-\Bigr]=\lfloor 2J_3 \rfloor, \quad
\Bigl[J_3,J_\pm\Bigr]=\pm J_\pm.
\label{2.27}
\end{equation}
We  define  $\lfloor r\rfloor \equiv \sin (h r)/\sinh$
in general. The fusion and braiding of the $\xi$ fields are
covariant under the quantum group action
\begin{equation}
 J_3\, \xi_{M}^{(J)}=M  \xi_{M}^{(J)},
\quad J_\pm\, \xi_{M}^{(J)}= \sqrt{\lfloor J \mp
M\rfloor
\lfloor J \pm M+1 \rfloor }\, \xi_{M\pm 1}^{(J)}.
\label{2.28}
\end{equation}
Indeed, first  in the
 the operator-product Eq.\ref{2.23}, the term
$<\!\varpi _J,{\{\nu\}} \vert V ^{(J_1)}_{J_2-J}
\vert \varpi_{J_2}\! >$ is a quantum-group invariant since it does not
involves the indices $M_1$, or $M_2$. Thus, by the basic property
of the Clebsch-Gordan coefficients, it follows that
 $\xi_{M_1}^{(J_1)}\>\xi_{M_2}^{(J_2)}$ also
gives a representation of the quantum group algebra \ref{2.27}
with  the co-product
generators
\beq
 {\Lambda_\pm}= J_\pm\otimes e^{ihJ_3}+e^{-ihJ_3}\otimes J_\pm, \quad
{\Lambda_3} = J_3\otimes 1+1\otimes J_3.
\label{2.29}
\eeq
The tensor product is defined so that
\beq
(A\otimes B)
\Bigl(\xi_{M_1}^{(J_1)}(\sigma)\>\xi_{M_2}^{(J_2)}(\sigma')\Bigr):=\bigl(A\xi_{M_1}^{(J_1)}(\sigma)\bigr)
\>\bigl(B\xi_{M_2}^{(J_2)}(\sigma')\bigr),
\label{2.30}
\eeq
Roughly speaking the operator-product correspond to ``adding''
the quantum group spins. Since the result of the fusion
should not depend upon the ordering, the non-commuativity
of the $\xi$ fields as quantum operators should precisely cancell
the lack of symmetry of the co-product, and Clebsch-Gordan coefficients.
This is why the braiding relations are given by the unversal
R-matrix. Thus {\bf the mathematical deformation of $sl(2)$, is
precisely governed by the truly quantum mechnical effects of the
theory}.
Second, since the universal $R$-matrix which gives the braiding
 commutes with the
co-product, the braiding relation Eq.\ref{2.24}
 is also covariant. This quantum-group symmetry, does not
prevent invariant coupling constants $g_{J_1 J_2}^{J_3}$
to appear\footnote{we
 simply see here an application of the well-known Wigner-Eckart
theorem, which is so much used, for instance,  in atomic physics}.
The value of these coupling constants makes the difference between
Liouville theory and the $SU(2)$ WZNW model at this level.

An other important feature  is that the fusion
and braiding matrices do not depend upon the Verma module to which
the fusion and braiding relations are applied. This is in contrast
with the usual MS expressions. In the latter case, this means that
the fusion and braiding matrices are not c-numbers since they  depend
upon the eigenvalue of the Liouville momentum. The corresponding
solution of the Yang-Baxter equation  thus takes a form
that differs from the solutions obtained from the R-matrix of
a quantum group.  It is only when (and if) this dependence
upon the Verma module may be removed that a direct connection with
quantum-group representations  is established. Note that, for
$A_n$-W-algebras ($n\not= 2$),  the same procedure works
but gives\cite{CG1} a deformation of $sl(n)$ that
differs from the standard one.
\subsection{The Liouville field, and matter-gravity couplings}
We consider closed surfaces following  ref.\cite{G5}.
The above  $\xi$ fields
together with their antiholomorphic counterparts $\bar \xi$ allow
us to reconstruct the fields
$\exp { -J\alpha_-\Phi(\sigma, \tau )}$. Imposing locality, and
conservation of the winding number determines this operator
completely. One finds
\begin{equation}
e^{\textstyle -J\alpha_-\Phi(\sigma, \tau )}=
{1\over \sqrt{\varpi}}\sum _{M=-J}^J\> (-1)^{J-M}  \>e^{ih(J-M)}\>
\xi_M^{(J)}(x_+)\,
{\overline \xi_{-M}^{(J)}}(x_-) \sqrt{\varpi}
\label{2.31}
\end{equation}
The quantum-group action on the   $\bar \xi$ field
is given by:
\begin{equation}
 {\overline J}_3\,{\overline \xi_{M}^{(J)}}=
M {\overline \xi_{M}^{(J)}},
\quad {\overline J}_\pm\,{\overline \xi_{M}^{(J)}}=
\sqrt{\lfloor J \mp
M\rfloor
\lfloor J \pm M+1 \rfloor }\,{\overline \xi_{M\pm 1}^{(J)}}.
\label{2.32}
\end{equation}
Thus if we define
\begin{equation}
 {\cal J}_\pm = J_\pm e^{-ih{\overline J}_3}+e^{ihJ_3}
{\overline J}_\pm, \quad
 {\cal J}_3 = J_3 + {\overline J}_3,
\label{2.33}
\end{equation}
we obtain  a representation of the quantum-group
algebra Eq.\ref{2.28}. Then one easily
checks that
\begin{equation}
{\cal J}_\pm \exp(-J\alpha_- \Phi)=
{\cal J}_3 \exp(-J\alpha_- \Phi)=0,
\label{2.34}
\end{equation}
so that the quantized Liouville field
is a quantum-group invariant. This is the quantized version of
Eq.\ref{2.11}. Thus the $sl(2)$ symmetry recalled in section 2
has been deformed by the quantization in the mathematically
standard way. The most general Liouville field
$\exp \bigl(-(J\alpha_-+\Jhat\alpha_+)\Phi\bigr)$, is given by
a similar expression in terms of the fields $\xi$, $\xihat$,
and $\bar \xi$, and $\bar \xihat$ (see ref\cite{G5} for details).

Next we cansider the dressing of minimal models with central charge
$D$ by Liouville with central charge $C$. The balance of central
charge requires
\begin{equation}
C+D=26.
\label{2.35}
\end{equation}
 We shall be concerned with the case
$D< 1$, where the Liouville theory is in its weakly coupled
regime $C>25$.  The quantum structure of the Liouville theory
just recalled
 is basically  a consequence of  operator differential
equations which   are equivalent to the Virasoro
Ward-identities that describe the decoupling of null vectors. Thus
the same operator algebra,
 with appropriate quantum deformation parameters also describes
the matter with $D<1$. We will thus have another copy of
the
quantum-group structure recalled above.
It will be distinguished by primes.
Thus we let
\begin{equation}
D=1+6({h'\over\pi}+{\pi\over h'}+2)=
1+6({\hhat'\over\pi}+{\pi\over\hhat'}+2),
\quad \hbox{with} \quad h'\hhat'=\pi^2,
\label{2.36}
\end{equation}
 The appropriate dressing by gravity is such that one is
concerned with matrix elements of  operators of the type
\begin{equation}
{\cal V}_{J',\,\Jhat'}(\sigma,\tau)\equiv
 e^{\textstyle ((\Jhat'+1)\alpha_--J' \alpha_+)\Phi(\sigma, \tau )}\>
e^{\textstyle -(J'\alpha'_-+\Jhat' \alpha'_+)
X(\sigma, \tau )},
\label{2.37}
\end{equation}
where $X$ is the matter field that is similar to the Liouville field.
In ref.\cite{G5}, the present approach was used to
determine the   three-point gravity-matter couplings:
\begin{equation}
\bigl <
\prod_{\ell=1}^3 {\cal V}_{J'_\ell,\,\Jhat'_\ell}(z_\ell,\,z^*_\ell)
 \bigr > =
{\cal C}_{1,2,3}\Bigr /
\Bigl (\prod_{k,l} \vert z_k-z_l \vert^2 \Bigr  ).
\label{2.38}
\end{equation}
For positiv $J'$, Eq.\ref{2.37} involve negative quantum-group spins.
This difficulty is equivalent with the need for negative number of
dressing operator that is encountered in the Coulomb gas approach.
Remarkably, this problem is precisely solved by symmetry $J\to -J-1$.
which was found in refs.\cite{G2,G3}. Finally the result is very simple
\beq
{\cal C}_{1,2,3}=
\prod_l {B_{J'_l,\,\Jhat'_l}\over
\left [
\Gamma\Bigl(1+2\Jhat_l+(1+2J_l)h/\pi\Bigr)
\right  ]^2}.
\label{2.39}
\eeq
It factorises, and thus gives back results of matrix models for
ratios of correlators that do not depend upon the normalizations.
The outcome of ref.\cite{G5} is that the symmetry
between
quantum-group spins $J$ and $-J-1$  is the explanation of the continuation in
the number of
screening operators discovered by Goulian and Li.
Moreover, and contrary to the previous discussions of this problem, the present
approach clearly separates the emission operators for each leg.
This  clarifies the structure of the dressing by gravity. It is shown,
in particular that the end points are not treated on the same footing
as the mid point. Since the outcome is completely symmetric this
suggests  the possibility of a picture-changing mechanism.

Before  leaving this topic we quote an older result\cite{CG2}
concerning open surfaces. The equivalent of Eq.\ref{2.31} is
\beq
e^{-\textstyle \alpha_- J\Phi(\sigma)}= \Bigl({h\over 16\pi^3}\Bigr)^J
\sum _{M,\,N}<J,M| A | J,N>
\xi_M^{(J)}(\sigma)\,\xi_N^{(J)}(2\pi- \sigma)
\label{2.40}
\eeq
where
\beq
A=  e^{-ihJ_3^2}\sum _{r,s=0}^\infty
e^{ih(r+s)J_3}{(J_+)^{r+s} e^{ihrs/2}\over
\lfloor r \rfloor \! !\, \lfloor s \rfloor \! !}.
\label{2.41}
\eeq
The matrix $A$ is the universal reflection matrix associated with
$U_q(sl(2))$, a concept that is gradually recognized as very important
for integrable systems with boundaries.
\section{BLACK HOLES  FROM TODA THEORIES}

In the present section, as well as in the coming ones, we restrict
ourselves to classical Toda field theories, that describe the classical
limits of conformal theories. Although this is a much simpler
problem than the quantum one, it already teaches  us a
lot, since the appropriate
operator quantization\cite{BG1,CG2} does preserve
the key features of the
classical Toda theories.
The basic point of these  theories
 is that
their dynamical equations, no matter how complicated, are derivable
from  flatness  conditions
\begin{equation}
\label{3.1}
\Bigl[\partial_+-{\cal A}_+,
\partial_--{\cal A}_-\Bigr]\equiv \partial_-{\cal A}_+
-\partial_+{\cal A}_-+\Bigl[ {\cal A}_+,{\cal A}_- \Bigr]=0,
\end{equation}
 which
allow  to obtain the general solutions in closed
form\cite{LS1}\cite{LS2}. We use $\partial_+$, and $\partial_-$
as short hands for $\partial/\partial x_+$, and
$\partial/\partial x_-$, respectively. The Lax pair
${\cal A}_\pm$ is systematically constructed once the
Lie algebra, the gradation, and the grading spectrum
of ${\cal A}_\pm$  are chosen\cite{LS1}\cite{LS2}. Consider a
finite-dimensional Lie group ${\bf G}$,
with Lie algebra $\cal G$, and a gradation
${\cal G} = \oplus_{m \in
{\bf Z}} {\cal G}_m$, with  $[{\cal G}_m, {\cal G}_n ]
 \in {\cal G}_{m+n}$. There is a grading operator $H$, such that
$[H, {\cal G}_m ] =
 2m  {\cal G}_{m}$.
The basic idea is to consider the flatness condition
Eq.\ref{3.1} taking ${\cal A}_\pm
\in {\cal G}_0\oplus {\cal G}_{\pm 1}$.
The (abelian) $A_N$-Toda theories considered in
the coming sections are associated
with the principal grading of $A_N$ where   $ {\cal G}_0$  is
the Cartan subalgebra, and where ${\cal G}_{\pm 1}$ is generated by the
simple (positive, resp. negative) roots.  It is derivable from
the Lagrangien
\begin{equation}
\label{3.2}
- {\cal L} = \frac{1}{2}{\sum^{N}_{a,b=1}}
K^{({A_N})}_{ab} \partial_+
\phi^{a}
\partial_- \phi^{b} + \sum_{a=1}^N
\prod_{b=1}^N   \exp \left(K^{({A_N})}_{ab} \phi^b\right ),
\end{equation}
The matrix $K^{({A_N})}_{ab}$ is the Cartan matrix of $A_N$.
The fields $\phi^a(x_+, x_-)$ are the bosonic Toda fields.
We shall come back to this model in the next section. At the
present time let us simply note that the fields $\phi^a$
 may be regarded
as components of a string-position in a $N$-dimensional
target space with a  metric  $G_{ab}=K^{({A_N})}_{ab}$.
It is somewhat trivial from this view point since this metric is
constant. This feature is no more true for non-abelian Toda theories,
namely those for which the subalgebra $\cal G_0$ is non-abelian.
This  was shown in ref\cite{GS}  in the
examples\footnote{Note that  there  is
no useful non-abelian version of the Toda lattice for the series
$A_N$}  of the Lie
algebras  $B_{N-1}$.
Associated   with  a  maximal  subalgebra  $D_{N-1}$  (of
maximal rank) in the Lie algebra $B_{N-1}$,
 we have the Lagrangian
\begin{eqnarray}
-{\cal L} =
\frac{1}{2}\Bigl[\sum_{j ,k =1}^{N-1} K_{jk}^{(D_{N-1})}
\partial_+
\phi^{j}
\partial_- \phi^{k}
& - & \tanh^2 \left ({\phi^{N-1} - \phi^{N-2}\over 2}\right  )
\frac{\partial \phi^N}{\partial z_+}
\frac{\partial \phi^N}{\partial z_-}\Bigr] \nonumber\\
& + & \sum_{i=1}^{N-1}
\prod_{j=1}^{N-1} \exp \left(K^{( D_{N-1})}_{ij} \phi^j\right ).
\label{3.3}
\end{eqnarray}
 In this case, the zero-grading part
is given by $ {\cal G}_0= gl(1)\oplus \cdots \oplus gl(1)
\oplus A_1$,
where the one-dimensional linear algebra
 $gl(1)$ appears $N-1$ times.
The $\phi$-dependent  part of the metric involves  the
 hyperbolic-tangent-square function  which is familiar
in the 2D black-hole game. In particular if we choose
$N=3$, we  find
\begin{equation}
\label{3.4}
-{\cal L } = \frac{1}{2}\left[2\sum_{i =1}^{2}
\partial_+
\phi^{i}
\partial_- \phi^{i}
 -  \tanh^2 \left ({\phi^{1} - \phi^{2}\over 2}\right)
\partial_+
\phi^{3}
\partial_- \phi^{3}
\right]
 +  \sum_{i =1}^{2} \exp (2\phi^i)
\end{equation}
In order to clarify the black-hole aspect,
it is convenient to change field variables. Let
\begin{equation}
\label{3.5}
\Phi=(\phi^1+\phi^2)/2,\quad
r=(\phi^1-\phi^2)/2, \quad
\theta=\phi^3/2.
\end{equation}
One gets
\begin{equation}
\label{3.6}
- {\cal L} =
\partial_+
\Phi
\partial_- \Phi
+\partial_+
r
\partial_-r
 -  \tanh^2 \left (r\right)
\partial_+
\theta
\partial_- \theta
 \nonumber\\
 +  \cosh (2r) e^{2\phi}.
\end{equation}
The target-space metric is
\begin{equation}
\label{3.7}
G =\left (\begin{array}{ccc}
1&0&0\nonumber\\
0&1&0\nonumber \\
0&0&-\tanh^2 r
\end{array}\right )
\end{equation}
One sees that
the first component,   which corresponds to the $\phi$ variable,
defines a subspace that
 decouples from the rest
from the viewpoint of Riemannian geometry.
In the ($r$, $\theta$)  space we have
$G=\left (\begin{array}{cc}
1&0\nonumber \\
0&-\tanh^2 r
\end{array}\right ),
$
which coincides  with Witten's black-hole metric\cite{W} exactly.

The outcome of ref.\cite{GS} is that non-abelian Toda
theories  provide exactly
solvable conformal systems in the presence of a
black hole. They correspond to gauged WZNW models where the
gauge group is nilpotent, and are thus basically different
  from the ones
currently considered, following Witten. The non-abelian
Toda potential  gives a cosmological term which may be integrated
exactly at the classical level.

\section{GEOMETRY OF CHIRAL SURFACES}

\subsection{The background}

The present discussion is concerned with the generalization
of the situation.
described in section 2  for 2D gravity.
On the one hand, the above Liouville dynamics is a
particular case of the
Toda dynamics\footnote{from now on we only consider the
principal grading}, which, as just recalled,
 exists for each simple Lie algebra. The Liouville
case is associated with $A_1$.
It was shown in ref.\cite{BG1} that general Toda
systems are related, through Noether theorem
to the non-linear extensions
of the Virasoro algebra called W-algebras. Thus going from
Liouville to general Toda should correspond, in general cordinates
to going from Einstein gravity to W gravity. As a first step towards
formulating these fascinating theories, we
next unravell  the geometrical meaning of
Toda systems. This is, at least  a way to see the geometrical meaning
of W algebras. We will show  that
 it corresponds to the extrinsic geometry of embeddings of
special (W) surfaces, a viewpoint which is   natural since in general,
we expect that conformal systems
and their W
generalizations are  to be connected with string theories.
The material covered untill the end is a summary of
 refs.\cite{GM1,GM2}.

\subsection{Chiral embeddings}
The basic objects are two-dimensional surfaces
embedded in K\"ahler manifolds.
We shall only consider $C^n$ here explicitly.  The case of
$CP^n$ is treated by using homogeneous coordinates in
$C^{n+1}$. We call $z$ and $\zb$ the surface
parameters. One may think of $z$ as an ordinary complex number, in
which case the parametrization is Euclidian, and $\zb$ is the complex
conjugate of $z$; or take $z$, and $\zb$ to be real, in which case
$x_0\equiv(z+\zb)/2$, and $x_1\equiv(z-\zb)/2$ are coordinates of the
Minkowsky type. The adjective chiral means function of a single variable
$z$ or $\zb$ (if $z$ is a complex variable this is
equivalent to analytic
or anti-analytic).
   We parametrize $C^n$ by coordinates
 $X^A$, $\Xb^{\Ab}$,
$1\leq A,\, \Ab \leq  n$.
A \souligne {chiral embedding} is defined by equations
of the form
\begin{equation}
X^A=f^A(z), \, A=1,\, \cdots,\, n, \quad
\Xb^{\Ab}=\fb^{\Ab}(\zb), \, \Ab=1,\, \cdots,\, n.
\end{equation}
where $f$ and $\fb$ are arbitrary f unctions. We call a
\souligne{$W$-surface}
 the
corresponding manifold $\Sigma$. We shall see that
its extrinsic geometry is directly related to $W$ transformations.
  It is convenient to introduce
the  matrix of inner products:
\begin{equation}
 g_{i \jb}\equiv \sum_{A}
\>\> f^{(i)A}(z)\>
  \fb^{(\jb)A}(\zb),
\quad
1\leq i,j \leq n.
\end{equation}
We use
$\partial $ and $\partialb$ as  short hands for
$\partial/\partial z$ and $\partial /\partial \zb$ respectively.
$f^{(i)A}$ stands for $(\partial)^i f^A$, and
$\fb^{({\jb})A}$ stands for $(\partialb)^{\jb} \fb^A$. Later on we shall
exhibit
a particular parametrization of $C^n$, called  W-parametrization,
  where the
 derivatives with $i$ or $\jb$ larger than one  will be replaced by
first order derivatives in other variables,
so that the covariance properties  of the present
discussion will become more transparent.
This section is    concerned with generic regular
points of $\Sigma$ where the Taylor expansions
of $f^A$ and $\fb^{\Ab}$ give linearly independent
vectors.  Then  ${\bf f}^{(a)}$,
and $\bbff^{(a)}$, $a=1$, $\cdots $, $n$,
 span the following

\begin{definition}{\bf Moving frame in $C^n$.}
Consider the vectors $\bfe_a$, and $\bbfe_a$, $a=1, \cdots, n$,
with components
\begin{equation}
 e_a^A = {1\over \sqrt{\Delta_a \Delta_{a-1}}}
\left | \begin{array}{ccc}
g_{1 \1b} & \cdots  & g_{a \1b} \\
\vdots    &         & \vdots  \\
g_{1 {\overline {a-1}} } & \cdots  & g_{a {\overline {a-1}}} \\
f^{(1)A} & \cdots &  f^{(a)A}
\end{array}
\right |, \qquad  e_a^{\Ab}=0,
\end{equation}
\begin{equation}
\eb_a^A=0,\qquad
\eb_a^{\Ab} = {1\over \sqrt{\Delta_a \Delta_{a-1}}}
\left | \begin{array}{ccc}
g_{\1b 1} & \cdots  & g_{\ab 1} \\
\vdots    &         & \vdots  \\
g_{\1b a-1} & \cdots  & g_{{\overline {a}} a-1} \\
\fb^{(1)\Ab} & \cdots & \fb^{(a)\Ab}
\end{array} \right |,
\end{equation}
 $\Delta_a$ is  the determinant
\begin{equation}
\Delta_a\equiv \left | \begin{array}{ccc}
g_{1 \1b} & \cdots  & g_{a \1b} \\
\vdots    &         & \vdots  \\
g_{1 \ab} & \cdots  & g_{a \ab}
\end{array}
\right |.
\end{equation}
\end{definition}
One may verify that the moving frame defined above is
orthonormal, that is,
\begin{equation}
( \bfe_a, \bfe_b)=( \bbfe_a, \bbfe_b)=0,\quad
( \bfe_a, \bbfe_b) =\delta_{a,b}.
\end{equation}
The vectors
${\bfe}_1$ and  ${\bbfe}_1$ are tangents to the surface,
while the other vectors are clearly normals. Thus the Gauss-Codazzi
equations may be  derived  by studying their  derivatives
along the W-surface $\Sigma$. One derives equations of the form
\beq
\partial \bfe_a=\sum_b\omega_{z a}^{\>b} \bfe_b, \quad
\partialb \bfe_a=\sum_b\omega_{\zb a}^{\>b} \bfe_b,
\quad
\partialb \bbfe_a=\sum_b\omegab_{\zb a}^{\>b} \bbfe_b, \quad
\partial \bbfe_a=\sum_b\omegab_{z a}^{\>b} \bbfe_b.
\end{equation}
which may be regarded as generalized Frenet-Serret formulae.
Next we recall the
\begin{definition}{\bf $\CP$ target space.}
The complex projective space $C^n$ is defined to be the
quotient of the space $C^{n+1}$  by
the equivalence relation
\beq
\bfX\sim {\bf Y},
\qquad {\rm if}\quad X^A = Y^A\rho(Y), \quad
 {\rm and }
\quad \Xb^{\Ab} = \bar{Y}^{\Ab}\rhob(\Yb).
\eeq
where $\rho$ and $\rhob$ are arbitrary  functions of
$n+1$ variables.
\end{definition}
Thus, the modification to go from $C^n$ to $CP^n$
is to use $n+1$ coordinates, so that now $A$, and $\Ab$
run from $0$ to $n$, and to write formulae that are covariant
under rescaling. This is achieved by letting the indices of the
matrix $g_{i\jb}$ run from $0$ to $n$ in the
$CP^n$ formulae for the moving frame. for this one includes
derivatives  of order $0$. Our basic result  is the following:
\begin{theorem}{\bf Gauss-Codazzi equations.}

Define Toda fields by
\beq
\phi_a=-\ln (\tau_a) , a=1,\, \cdots,\, n; \>
\tau_a\equiv \left | \begin{array}{ccc}
g_{0 0} & \cdots  & g_{a 0} \\
\vdots    &         & \vdots  \\
g_{0 a} & \cdots  & g_{a a}
\end{array}
\right |.
\eeq
The integrability conditions of the Frenet-Serret equations
for the embedding in $CP^n$ ( analogous to
Eqs.\ref{2.7})  coincide with the Toda equations associated
with $A_n$:
\beq
\partial \partialb   \phi_i+
\exp \left ( \sum_{j=1}^{n} K_{i j} \phi_j \right )
=0.
\eeq
\end{theorem}
$K$ is the Cartan matrix associated with $A_n$.
The functions $\tau_a$ relevant for  $CP^n$ are similar to the
$\Delta_a$ of Eq.\ref{2.5}. except that they include the
derivatives of zeroth order.

In conclusion: The $A_n$ Toda dynamics describes the extrinsic
geometry of W surfaces.

\subsection{Some basic facts about $A_n$ Toda theories}
Its general solution is of the form
\begin{equation}
e^{-\phi_k}=\sum_{i_1< \cdots <  i_k}
\left | \begin{array}{ccc}
\chi^{i_1}  & \cdots  & \chi^{ i_k} \\
\chi^{(1)\, i_1}  & \cdots  & \chi^{(1)\, i_k} \\
\vdots     & \cdots         & \vdots  \\
\chi^{(k-1)\, i_1}  & \cdots  & \chi^{(k-1)\, i_k} \\
\end{array}
\right |
\left | \begin{array}{ccc}
\chib^{ i_1}  & \cdots  & \chib^{ i_k} \\
\chib^{(1)\, i_1}  & \cdots  & \chib^{(1)\, i_k} \\
\vdots     & \cdots         & \vdots  \\
\chib^{(k-1)\, i_1}  & \cdots  &
\chib^{(k-1)\, i_k} \\
\end{array}
\right |.
\end{equation}
where $k$ runs from $1$ to $n$. It is expressed in terms of
$n$ functions of $z$ $\chi^1,\, \cdots,\, \chi^n$, and
$n$ functions of $\zb$ $\chib^1,\, \cdots,\, \chib^n$.
Upper indices in between parenthesis denotes derivatives.
These n functions $\chi^k$ (resp. $\chib^k$) are restricted to be
solution of the same differential  equation $\chi^{(n+1) k}
-\sum_{\ell=0}^{n-1}  U_{n+1-\ell} \chi^{(\ell) k} =0$
(resp. $\chib^{(n+1) k}
-\sum_{\ell=0}^{n-1} \Ub_{n+1-\ell} \chib^{(\ell) k} =0$). The set of
potentials
$\{U_{\ell}, \ell=2,\, \cdots n+1\}$ and
$\{\Ub_{\ell}, \ell=2,\, \cdots n+1\}$ , each generate
a realization of the
$A_n$ W algebra by Poisson brackets. These are non-linear
generalizations of the Virasoro algebra (conformal transformations).
 The Toda dynamics is non-chiral,
and this is why the W algebra appears twice
(for the holomorphic and anti-holomorphic components).

It follows that from  the above geometrical derivation  of the Toda
eqautions, we may discuss the geometrical
meaning of the W transformations.

\subsection{Connection with the WZNW models}
It is known, in general, that there is one W algebra associated with each
 simple\footnote{The  non-simple Lie algebras  are trivially
 reduced to the simple case by separating the invariant subalgebras}
  Lie algebra $\cal G$. This appears in several ways.
First, as we have already seen, there is a Toda theory, and, hence,
two PB realizations of the associated W-algebra for any given $\cal G$.
On the other hand, consider the affine Lie algebra $\widetilde {\cal G}$
associated with $\cal G$.   The associated non-chiral theory is the
WZNW model whose quantum solutions are  given by representations of
$\widetilde {\cal G}$. It is possible to derive the Toda theory
from the WZNW model by conformal reduction\cite{Dublin}.
Here we have the
\begin{theorem}{\bf Gauss decomposition from moving frame.}
The moving-frame equations may be written as
\begin{equation}
\bfe_a=\sum_{b\leq a} C_{ab}(z, \zb) \>
\sqrt{{\tau_{a}\over \tau_{a+1}}} \> \bff^{(b)}(z), \quad
\hbox{with}\> C_{aa}=1,
\end{equation}
\begin{equation}
\bbfe_a=\sum_{b\leq a} A_{ba}(z,\zb)\>
\sqrt{{\tau_{a}\over \tau_{a+1}}}\>
\bbff^{(b)}(\zb), \quad\hbox{with}\> A_{aa}=1.
\end{equation}
The matrix  $\theta=g^{-1}$ s such that

\noindent 1) It has  the Gauss decomposition
\beq
\theta  = A B C,
\end{equation}
where  $A$, and $C$, which appear in eqs.\ref{2.11}, and \ref{2.12}
are triangular with diagonal elements equal to one, and
$$
B=\exp\left ( \sum_i h_i \phi_{i+1}\right ).
$$
$h_i$ are the Cartan generators.
\noindent 2) It is a solution of the conformally reduced
WZNW model associated with $A_n$.
\end{theorem}
\section{Geometry of Toda hierarchy}

The Toda equations are a subsystem of the
Toda hierarchy\cite{DJKM}. (This is the
non-chiral version of the fact that  the Virasoro algebra
is identical with the second Poisson bracket
 of KdV, and that W algebras are obtained from
KP hierarchies and Gelfand Dicki brackets). Introduce the additional
variables as coordinates in our geometrical
embedding problem. This is best
done using the free-fermion formalism. Let
\begin{eqnarray}
\left[ \psi_n, \psi_m \right]_+ & =
& \left[\psi^+_n, \psi^+_m\right]_+ = 0,\nonumber\\
\left[\psi_n, \psi^+_m\right]_+ & = & \delta_{n,m},
\qquad (~n,m~=~0,~1,~\cdots)
\\
\psi_n\ket{\emptyset}  =  0, &\qquad&
\bra{\emptyset}\psis_n  =  0  \qquad \forall n.
\end{eqnarray}
We use the semi-infinite indices $n=0,1,2,\cdots, \infty$ for
the fermion-operators.
The vacuum states $\ket{\emptyset}$ and $\bra{\emptyset}$
correspond to the
no-particle states.  The $n$--particle ground state is
 created from
them in the standard way:
\begin{equation}
\ket{n}  =  \psis_{n-1}\psis_{n-2}\cdots\psis_0\ket{\emptyset},
\quad
\bra{n}  =  \bra{\emptyset}
\psi_0\psi_1\cdots\psi_{n-1}.
\end{equation}
The  current operators,
\begin{equation}
J_n = \sum_{s=0}^\infty \psis_{n+s}\psi_s,  \qquad
\bar{J}_n = \sum_{s=0}^\infty \psis_s \psi_{n+s},
\label{current}
\end{equation}
will be taken as  Hamiltonians as one  does  for  the
KP hierarchy.
The r\^ole   of these fermions may be understood as follows.
Take the case where $z$ is a complex variable. Then
the embedding functions $f^A$ are analytic, and  each  of them is
entirely determined
by its  Taylor expansion around a single point of its analyticity
domain.  Its behaviour at any other point of its  Riemann surface is
fixed by analytic continuation. The following  free-fermion formalism
realizes this continuation automatically. Consider the Taylor
 expansions
at the point $z$:
\beq
f^A(z+x) = \sum_{s=0}^\infty
f^{(s)A}\! (z)\, {x^s\over s!},\quad
\fb^{\Ab}(\zb+{\bar x}) = \sum_{s=0}^\infty
\fb^{(s)A} \! (\zb)\,
{{\bar x}^s\over s!}.
\eeq
To these developements,  we associate the free-fermion operators,
\beq
\psifz{A}{z} = \sum_{s=0}^\infty f^{(s)A}\! (z)\> \psi_s ,\quad
\psifsz{\Ab}{\zb} = \sum_{s=0}^\infty \fb^{(s)A}\!  (\zb)\>
\psis_s.
\eeq

The basic property of these  operators are
\begin{proposition}{\bf Fermionic representation of chiral
functions.}

\noindent 1) Any change of the Taylor-expansion point $z$, $\zb$
 can be absorbed by
the action of the Hamiltonians $J_1$, and $\Jb_1$.
In particular, one has
\beq
\psifz{A}{z}=e^{-J_1z}\>\psifz{A}{0}\> e^{J_1z}\quad
\psifsz{\Ab}{\zb}=e^{\Jb_1\zb}\>\psifsz{\Ab}{0}\> e^{-\Jb_1\zb}.
\eeq
2) The embedding functions are represented by the fermion
 expectation-values
\beq
\label{ferm}
f^A(z) = \bra{\emptyset}\psifz{A}{z_0} e^{J_1(z-z_0)}\ket{1},\qquad
\fb^{\Ab}(\zb) = \bra{1}e^{\Jb_1(\zb-\zb_0)} \psifsz{\Ab}{\zb_0}
\ket{\emptyset}.
\eeq
\end{proposition}
\begin{definition}{\bf KP-parametrization of $CP^{n}$. }

Given a chiral surface embedded into $CP^n$,
the associated KP-parameters of
the target space are
$n+1$ variables $z^{(0)}$, $z^{(1)}=z$, $z^{(2)}$, $\cdots$,
$z^{(n)}$,
noted  $[z]$, and $n+1$ variables $\zb^{(0)}$, $\zb^{(1)}=\zb$,
$ \zb^{(2)}$, $\cdots$, $\zb^{(n)}$,
noted  $[\zb]$.  The change of coordinates from $X^A$, $\Xb^{\Ab}$
to $[z]$,  $[\zb]$ is defined by
\beq
X^A=f^{A}([z]),\qquad \Xb^{\Ab}=\fb^{\Ab}([\zb])
\eeq
where
 $f^{A}([z])$, and
 $\fb^{\Ab}([\zb])$, are  the solutions of the
equations
\beq
\label{diff}
\frac{\partial}{\partial z^{(\ell)}} f^A([z])
= \frac{\partial^\ell}{\partial z^\ell} f^A([z]),
\quad
\frac{\partialb}{\partial \zb^{(\ell)}} \fb^{\Ab}([\zb])
= \frac{\partialb^\ell}{\partial \zb^\ell} \fb^{\Ab}([\zb])
\eeq
with the initial conditions $f^A([z])= f^A(z)$
for  $z^{(0)},\ z^{(2)}, \cdots, z^{(n)}=0$, and
$\fb^{\Ab}([\zb])= \fb^{\Ab}(\zb)$
for  $\zb^{(0)},\ \zb^{(2)}, \cdots, \zb^{(n)}=0$.
\end{definition}
These coordinates  coincide with  the higher variables
of the KP hierarchy. Indeed,  their definition  is
most natural in  the free-fermion language,
where it is easy to see that
\beq
f^A([z]) = \bra{\emptyset}\psif{A}\eJz\ket{1},\quad
\fb^{\Ab}([\zb]) = \bra{1}\eJzb\psifs{\Ab}\ket{\emptyset}.
\eeq
The dependence in $[z]$ and $[\zb]$ is
dictated by the action
of the higher currents $J$, $\bar{J}$,
defined by Eq.\ref{current},that is,
$J_1z\to \sum_{i=0}^n J_iz^{(i)}$,
$\Jb_1\zb\to \sum_{i=0}^n \Jb_i\zb^{(i)}$ in Eq.\ref{ferm}.
The basic virtue of the
KP coordinates   is that they allow us to extend the previous discussion
away from the W surface (they parametrize at least a neighborhood of it) in
such a way that is becomes covariant. In particular,
the metric $g$ has the
following extension
\beq
g_{i\jb}([z],[\zb]) =
\sum_A\partial_if^A([z])\, \partialb_j\fb^{A}([z]),
\quad \partial_i\equiv \frac{\partial}{\partial z^{(i)}},
\quad\partialb_i\equiv \frac{\partialb}{\partialb \zb^{(i)}}.
\eeq
Now, only first-order  derivatives appear.
This
expression
coincides with the
 {\it true } Riemannian metric with respect to the KP
coordinates.
\subsection{$W$ transformations}
A general infinitesimal W-transformation is a change of
embedding functions which takes the form
\beq
\delta_W f^A(z)=\sum_{j=0}^n w^j(z) \partial^{(j)} f^A(z),
\quad \delta_W \fb^{\Ab}(\zb)=\sum_{j=0}^n
\bar w^j(\zb) \partial^{(j)} \fb^{\Ab}(\zb),
\eeq
It is not difficult to show there exists a  unique extension such that
the differential equation \ref{diff} is left invariant.
It  is of the form
\beqa
\delta_W f^A([z])&=&\sum_r W^r([z]) \partial_r f^A([z]),\nonumber \\
\delta_W \fb^{\Ab}([\zb])&=&\sum_r \Wb^r([\zb]) \partial_r \fb^{\Ab}([\zb])
\eeqa
where
$W^r$ and $\Wb^r$  are functionals of $w^j$ and $\bar w^j$ respectively.
Only first order derivative appear. Thus the W transformation become
extended as reparame\-trizations
\beq
\delta_W z^{(r)}= W^r([z]), \quad
\delta_W \zb^{(r)}= \Wb^r([\zb]).
\eeq
They become particular diffeomorphisms of $CP^n$. Thus they are extended
as  linear transformations.
\subsection{Dynamical equations}
The followings topics are discussed in ref.\cite{GM2}

The above functions $\tau_a$ when extended become tau-functions
in the sense   of Miwa-Jimbo-Sato\cite{DJKM}.

The KP coordinates  are related with  a generalized
moving frame, whose integrability condition is equivalent to the
 the well-known Zakharov-Shabat of the $A_n$ Toda hierarchy.

The extension of the associated WZNW model gives solutions of a
 2n dimensional
generalization of the WZNW equations where the currents
are replaced by the
Christoffel symbols of the KP coordinates.

\section{SINGULAR POINTS, GLOBAL STRUCTURE}
At this point,  it is useful to change the viewpoint, and make use of
Grassmannians.   This  part is draws a lot of inspiration from
ref.\cite{GH}
\subsection{Associated mappings}
\begin{definition}{\bf  Associated mappings.}

Consider the family of osculating hyperplanes with contact
of order $k$ denoted ${\cal O}_k$
($ k=1,\cdots,n$) to the original W-surface.
With $CP^n$ as the target space, this family defines an
embedding into the Grassmannian $G_{n+1,k+1}$,
 which we  call  the
$k$th associated mappings to  the original W-surface.
\end{definition}
This  formulation
looks different, but  is equivalent to the construction
of the moving frame and  only uses the intrinsic geometries
of the induced metrics for $k=1,\cdots,n$.
In practice, what this means is that,  instead of
forming moving-frame vectors $\bfe_k$ out of
$\bff,\cdots, \bff^{(k)}$ ( $k=1,\cdots, n$), we consider the
nested
osculating planes ${\cal O}_1\subset{\cal O}_2\subset\cdots
\subset{\cal O}_n$.  It is obvious that those two have the same
information.
It is well-known that the Grassmannians are K\"ahler manifolds.
  The induced metric on the
$k$th associated surface
 in $G_{n+1,\, k+1}$ is simply,
\begin{equation}
\label{met}
g_{z\, \zb}^{(k)}=\partial \partialb \ln \tau_{k+1}(z,\zb),\quad
g_{z\, z}^{(k)}=g_{\zb\, \zb}^{(k)}=0,
\end{equation}
so that the Toda field $\phi_{k+1}\equiv
-\ln (\tau_{k+1})$ appears naturally.
Thus  $-\phi_{k+1}$ is equal to the  K\"ahler potential
of the $k$th associated surface.
At this point, it is very clear that by considering the associated
surfaces, we can restrict ourselves to  intrinsic geometries.
In the discussion of  section 4, the Toda equation
came out from the Gauss-Codazzi equation.
Here, it is equivalent to the local Pl\"ucker formula
\begin{equation}
\label{localP}
R_{z\, \zb}^{(k)} \sqrt{g_{z\, \zb}^{(k)}}
=-g_{z\, \zb}^{(k+1)}+2g_{z\, \zb}^{(k)}
-g_{z\, \zb}^{(k-1)}.
\end{equation}
where $R_{z\, \zb}^{(k)}$ is the only non-vanishing component of the
intrinsic Riemann tensor on the $k$th surface.

\subsection{The instanton-numbers of a W-surface}

A key point in the coming discussion is to use topological
quantities that are instanton-numbers.
 W-surfaces are instantons of the associated
non-linear $\sigma$-model. The general situation is as follows.
W-surfaces are characterized by their chiral parametrizations
which   thus satisfy
the Cauchy-Riemann relations. These are self-duality equations
so that the coordinates of a W-surface  define fields that
are solutions of the associated non-linear $\sigma$ model,
with an  action  equal to
the topological instanton-number.  For a general K\"ahler
manifold $M$ with coordinates $\xi^\mu$
and $\xib^{\mub}$, and
metric $h_{\mu \mub}$, the action associated with 2D manifolds
of $M$ with equations $\xi^\mu=\varphi^\mu(z,\,\zb)$,
and $\xib^{\mub}=\varphib^{\mub}(z,\, \zb)$ is
\begin{equation}
S={1\over 2} \int d^2 x\>  h_{\mu \mub} \> \partial_j  \varphi^\mu
\partial_j \varphib ^{\mub}.
\end{equation}
In this short digression  we let $z=x_1+i x_2$,
and $\partial_j=\partial/\partial x_j$.
The instanton-number is defined by
\begin{equation}
\label{Inumb}
Q={i\over 2\pi}\int d^2 x \> \epsilon_{jk}\>
h_{\mu \mub} \> \partial_j  \varphi^\mu
\partial_k \varphib ^{\mub}.
\end{equation}
For W-surfaces and their associated surfaces,
 $\partialb \varphi^{\mu}=\partial \varphib^{\mub}=0$, and
one has $S=\pi Q$. $Q$ is proportional to the integral
of the determinant of the induced metric, that is
$Q={i\over 2\pi}\int d^2 x \partial \partialb \ln \tau_1$.

Moreover we may also apply  formula Eq.\ref{Inumb}
to the $k$th associated surface. This gives
\begin{definition}{\bf Higher instanton-numbers of the W-surface.}

The
$k$th instanton number of the W-surface
$Q_{k+1}$ is defined by,
\begin{equation}
Q_{k+1}\equiv {i\over 2\pi} \int _{\Sigma}
dzd\zb g_{z\, \zb}^{(k)}, k=1,\cdots, n-1.
\end{equation}
\end{definition}
Its topological nature is obvious from Eq.\ref{met} which shows
that the integrand is indeed a total derivative. The collection
of the ($k$-th) instanton-numbers together with the original one
$Q\equiv Q_1$
gives a set of topological quantities
which characterize the global properties of the original W-surface.
\subsection{Global classification of W surfaces}
In   section 4, we have constructed the moving frames
at the point where the tau-functions
are regular.
 When those
functions become irregular,  we meet an obstruction
to derive the moving frames. In
the WZNW language, this  signals that
there appears a global obstruction to the Gauss decomposition.
Toda equations  should be modified at these points.
In the following, we study the structure of
such singularities.
\subsection{ Gauss-Bonnet Theorem for W-surfaces}
For isolated singularities\footnote{if there is a cut with a
finite number
of sheets, one takes the appropriate covering},
the obstruction to constructing the moving frame may be reduced to the
vanishing of certain terms in the Taylor expansion. The latter is
characterized by the
ramification indices $\beta_k$ which are integers.
Apply the  Gauss-Bonnet theorem for
each of the $k$th associated surfaces by computing
$\int_{\Sigma_\epsilon} R_{z\, \zb}^{(k)}
\sqrt{g_{z\, \zb}^{(k)}}$. The  integral is first computed over
$\Sigma_\epsilon$ where  small neighborhoods of singularities
are removed.
The ramification
indices at singularity was  defined so that at a singular
point the induced metric of the $k$th associated surface behaves as
\begin{equation}
g_{z\, \zb}^{(k)}\sim (z-z_0)^{\beta_k(z_0)}
 (\zb-\zb_0)^{\betab_k(\zb_0)}
\tilde g_{z\, \zb}^{(k)},
\end{equation}
where $\tilde g_{z\, \zb}^{(k)}$ is regular at $z_0$, $\zb_0$. Since we do not
assume that $\overline {f(z)}=\fb(\zb)$,  $\beta_k(z_0)$ and
$\betab_k(\zb_0)$ may be different. By letting $\epsilon\to 0$,
one sees, that the contribution of the singularities
to the Gauss-Bonnet formula is proportional
to the k-th ramification index
\begin{equation}
\beta_k\equiv {1\over 2}
\sum_{ (z_0, \zb_0)\in \Sigma}\left ( \beta_k(z_0)+
\betab_k(\zb_0)\right ).
\end{equation}
The contribution of the regular part does not depend upon k, since
changing k  is equivalent to
using a different complex structure, while the
result is equal to the Euler characteristic
that does not depend upon it.
The Gauss-Bonnet theorem for the
$k$th associated surface  finally gives
\begin{equation}
{i\over 2\pi} \int _{\Sigma} dzd\zb R_{z\, \zb}^{(k)}
\sqrt{ g_{z\, \zb}^{(k)}}
=2-2g+\beta_k.
\end{equation}
Combining these last relations with Eqs.\ref{localP},
one arrives at the
\begin{theorem}{\bf Global Pl\"ucker formulae}
The genus $g$ of a W-surface is related to its  instanton-numbers
and ramification-indices by the relations
\begin{equation}
2-2g+\beta_k=2Q_k-Q_{k+1}-Q_{k-1}, \quad \left \vert
\begin{array}{cc} k&=1,\cdots, n\nonumber \\
Q_{n+1}&\equiv 0, \>
Q_{0}\equiv 0 \end{array}\right.
\end{equation}
\end{theorem}
\subsection{Simple example}
Consider the case of Liouville theory, for which one has $n=1$.
The simplest chiral surface corresponds to
$$
f^0=1,\> f^1=z;\quad \fb^0=1,\> \fb^1=\zb; \quad
\tau_1=1+z\zb,\quad \tau_2=1.
$$
The instanton number is thus
$
Q=(i/ 2\pi) \int dz d\zb/ (1+z\zb)^2=1
$.
On the other hand, one has
$$R_{z\, \zb}
\sqrt{ g_{z\, \zb}}=-\partial\partialb \ln(g_{z\, \zb})=
{2\over (1+z\zb)^2}
$$
so that
$$
{i\over 2\pi} \int dz d\zb R_{z\, \zb}
\sqrt{ g_{z\, \zb}}=2 Q =2,
$$
and one verifies that the above formulae indeed hold with
vanishing genus and ramification index.  In this example, the
Liouville solution coincides with the metric of
 the Lobatchevki half-plane.
Upon quantization, it corresponds to the $SL(2,C)$ invariant
vacuum of the Liouville theory.


\begin{thebibliography}{99}

\bibitem{GM1} J.-L. Gervais, Y. Matsuo,
{\sl Phys. Lett.} {\bf B274} (1991) 309.

\bibitem{GM2} J.-L. Gervais, Y. Matsuo,
``$A_n$--W-geometry'' preprint LPTENS-95/35, hep-th 9201026,
Comm. in Math. Phys. to be published.



\bibitem{BG1}
A. Bilal and J.-L. Gervais, {\sl Phys. Lett.} {\bf B206}
(1988) 412; {\sl Nucl. Phys. } {\bf B314} (1989) 646;
{\sl Nucl. Phys.} {\bf B318} (1989) 579.



\bibitem{GN1} J.-L. Gervais, A. Neveu,  \np B199, 59,
1982.

\bibitem{GN2} J.-L. Gervais, A. Neveu,  \np B202, 125,
1982.

\bibitem{GN4} J.-L. Gervais, A. Neveu \np B238, 125,
1984;
\np B238, 396, 1984.



\bibitem{DJKM}
M. Sato, {\sl RIMS Kokyuroku} {\bf 439} (1981) 30;
E. Date, M. Jimbo, M.Kashiwara and T.Miwa,
``Transformation Groups for Soliton Equations" in {\it
Proc. of RIMS Symposium on Non-linear  Integrable Systems-
Classical Theory and Quantum Theory} (Kyoto, Japan, May 1981)
(World Scientific Publication Co. Singapore, 1983);
G.Segal and G.Wilson, {\sl Pub. Math. IHES} {\bf 61} (1985) 5.

\bibitem{DS}
V.G. Drinfeld and V.V. Sokolov, {\sl Journ. Sov. Math.}
{\bf 30} (1985) 1975.

\bibitem{Dublin}
P.Forg\'acs, A.Wipf, J.Balog, L.Feh\'er, and
L.O'Raifeartaigh, {\sl Phys. Lett.} {\bf 227B} (1989) 214;
J.Balog, L.Feh\'er,
L.O'Raifeartaigh, P.Forg\'acs and A.Wipf,
{\sl Ann. Phys.} (N.Y.) {\bf 203} (1990) 76;
{\sl Phys. Lett.} {\bf 244B} (1990) 435;
L.O'Raifeartaigh and A.Wipf,
{\sl Phys. Lett.} {\bf 251B} (1990) 361;
L.O'Raifeartaigh, P.Ruelle, I. Tsutsui and A.Wipf,
{\it W-Algebras for Generalized Toda Theories}
Dublin preprint DIAS-STP-91-03, {\sl Comm. Math. Phys.}
to appear;
L.Feh\'er, L.O'Raifeartaigh, P.Ruelle, I. Tsutsui and A.Wipf,
{\it Generalized Toda theories and W-algebras associated with integer
grading},
Dublin preprint DIAS-STP-91-17, {\sl Ann. Phys.} to appear;
L.O'Raifeartaigh, P.Ruelle and I. Tsutsui,
{\sl Phys. Lett.} {\bf 258B} (1990) 359.

\bibitem{GH} P. Griffiths, S. Harris, ``Principles of algebraic
geometry'' Wiley-Interscience (1978).

\bibitem{B} O. Babelon,
\pl   B215,  523, 1988.

\bibitem{G1} J.-L. Gervais,  \cmp 130, 257, 1990.

\bibitem{G2} J.-L. Gervais, \pl B243, 85, 1990.

\bibitem{G3} J.-L. Gervais, \cmp 138, 301, 1991.

\bibitem{G4} J.-L. Gervais, ``On the algebraic structure
of
Quantum gravity in two dimensions'',
  Proceedings of the
Trieste Conference on  {\sl Topological Methods in Quantum Field
Theories}, June 1990, W. Nahm, S. Randjbar-Daemi, E.
Sezgin, E. Witten editors, World Scientific.

\bibitem{CG1} E. Cremmer, J.-L. Gervais,
 \cmp 134, 619, 1990.

\bibitem{G5} J.-L. Gervais, ``Quantum group derivation
of 2D gravity-matter coupling'' Invited talk at
the Stony Brook meeting {\sl String and Symmetry 1991},
LPTENS preprint 91/22, Nucl. Phys. B to be published.

\bibitem{GN6} J.-L. Gervais, A. Neveu, \pl 151B, 271,
1985.

\bibitem{MS} G. Moore, N. Seiberg, \cmp  123, 177, 1989.

\bibitem{CGR} E. Cremmer, J.-L. Gervais, J.F. Roussel,
LPTENS preprint to be published.

\bibitem{CG2} E. Cremmer, J.-L. Gervais,
 \cmp 144, 279, 1992.

\bibitem{LS1}
A.N.Leznov, M.V.Saveliev: Acta Appl. Math. {\bf 16},1 (1989).
\bibitem{LS2}
A.N.Leznov, M.V.Saveliev: Lett. Math. Phys. {\bf 6},505 (1982);
Comm. Math.
Phys. {\bf 89},59 (1983).

\bibitem{GS} J.-L. Gervais, M. Saveliev,
\pl 274, 309, 1992.

\bibitem{W} E. Witten: Phys. Rev. {\bf D44}  (1991) 314,
{\sl On black holes in string theory};  Lecture at Strings '91,
Stonybrook, June 1991.




\end{thebibliography}
\end{document}